# Perspectives of Electrically generated spin currents in ferromagnetic materials

Angie Davidson[1*], Vivek P. Amin[2,3*], Wafa S. Aljuaid[1], Paul M Haney[3], Xin Fan[1]

1. Department of Physics and Astronomy, University of Denver, Denver, CO 80210, USA
2. Maryland Nanocenter, University of Maryland, College Park, MD 20742, USA
3. Physical Measurement Laboratory, National Institute of Standards and Technology, Gaithersburg, Maryland 20899, USA

**Abstract**

Spin-orbit coupling enables charge currents to give rise to spin currents and vice versa, which has applications in non-volatile magnetic memories, miniature microwave oscillators, thermoelectric converters and Terahertz devices. In the past two decades, a considerable amount of research has focused on electrical spin current generation in different types of nonmagnetic materials. However, electrical spin current generation in ferromagnetic materials has only recently been actively investigated. Due to the additional symmetry breaking by the magnetization, ferromagnetic materials generate spin currents with different orientations of spin direction from those observed in nonmagnetic materials. Studies centered on ferromagnets where spin-orbit coupling plays an important role in transport open new possibilities to generate and detect spin currents. We summarize recent developments on this subject and discuss unanswered questions in this emerging field.

paul.haney@nist.gov

xin.fan@du.edu

* These two authors contributed equally to the manuscript.

1. Introduction

The electrical generation of spin currents is at the heart of spintronics research. In a ferromagnet, an electric field can generate a spin-polarized current because majority and minority carriers have different conductivities[1]. The spin polarized current generated by a ferromagnetic layer can transfer its spin angular momentum to another ferromagnetic layer within the same heterostructure. This process, known as spin transfer torque [1, 2], does not require spin-orbit coupling; in fact, its earliest explanation relied on conservation of total spin and assumed vanishing spin-orbit coupling. Spin transfer torque enables current-driven magnetization switching and provides a write mechanism for state of the art magnetic random access memories, which are now produced on an industrial scale.

Electrical spin current generation can also be directly achieved via spin-orbit coupling in nonmagnetic materials. In 1971, Dyakonov and Perel [3] predicted the spin Hall effect, in which an unpolarized charge current gives rise to a spin current via the spin-orbit interaction in the bulk of a nonmagnetic material. Unlike the spin filtering effect in ferromagnets, where the charge current and spin flow are in the same direction, the spin Hall effect generates spin currents such that their spin flow and spin direction are orthogonal to each other and to the charge current direction. The spin Hall effect and its reciprocal effect, the inverse spin Hall effect, have been experimentally detected in nonmagnetic materials by numerous methods [4-9].

While the spin Hall effect arises from bulk spin-orbit coupling, other spin-to-charge conversion effects like the Rashba-Edelstein effect arise from interfacial spin-orbit coupling. The Rashba-Edelstein effect results in an electrically generated spin accumulation, but unlike the spin Hall effect, does not result in a non-equilibrium spin current. Additional mechanisms to electrically generate spin currents include spin swapping [10] and interface generated spin currents [11]. In nonmagnet/ferromagnet bilayers, the spin currents generated in the bulk layers and the spin accumulations generated at the interfaces can exert spin torques on the ferromagnetic layer. In this context, spin torques are often referred to as spin-orbit torques since they arise from spin-orbit coupling. Spin-orbit torques can be used to electrically switch the magnetization direction [12, 13], manipulate magnetic textures [14-18] and drive magnetization auto-oscillations [19-22]. Each of these effects possess a reciprocal partner described via Onsager relations. For example, the inverse spin Hall effect refers to the generation of an electric field from an injected spin current. The inverse Rashba-Edelstein effect and the inverse spin Hall effect have demonstrated exciting new applications such as spin-based thermal energy harvesting [23, 24] and Terahertz pulse generation [25, 26].

Until recently, direct electrical spin current generation via spin-orbit coupling have focused on nonmagnetic conductors. In these studies, ferromagnetic conductors are typically also present and serve as either a spin current source or spin current detector. However, more detailed studies of the electrical generation of spin currents in ferromagnets via spin-orbit coupling has only received attention in recent years, even though ferromagnets strongly manifest spin-orbit coupling as evidenced by the anomalous Hall effect and anisotropic magnetoresistance. It is commonly assumed that spin currents generated within ferromagnets have spin directions that are aligned with the magnetization. Spin currents of this form would be more constrained than those allowed in nonmagnets. However, the reality is completely the opposite: In general, electrical spin current generation in ferromagnets is *less* constrained than in nonmagnetic materials due to the additional symmetry breaking arising from the magnetization.

---

[1] *Note that the physical mechanisms required to generate spin polarized current can depend on boundary conditions. If the boundary condition is that the current entering the system is unpolarized, then spin-flip scattering is necessary for polarizing the current. If the boundary condition only specifies the electric field at the system edge, then spin-dependent conductivities are sufficient to obtain spin-polarized current.*

In this article, we explore the nature of spin currents in magnetic materials by distinguishing between the behaviors of spins oriented longitudinal or transverse to the magnetization. In addition to this distinction, spin current generation in ferromagnets could also be categorized as intrinsic (arising from perturbation of electronic wavefunctions) or extrinsic (impurity scattering-based) mechanisms, as well as bulk or interfacial. These categories are not mutually exclusive, and co-exist in most, if not all, cases. Further research is required to determine the most important mechanisms and simplest categorization of spin current generation in magnetic materials.

We first consider the behaviors of spins oriented longitudinal or transverse to the magnetization. While both longitudinal and transverse spins undergo spin relaxation via spin-orbit scattering, transverse spins may also precess around the magnetization direction due to exchange coupling. *Incoherent* spin precession destroys the net spin density transverse to the magnetization. This process is known as spin dephasing. Since total angular momentum is conserved, the lost spin angular momentum from the transverse spins is transferred to the magnetization (if spin-orbit scattering is weak), giving rise to the well-known spin transfer torque [1, 2]. This picture of spin dephasing applies to spin currents being injected into ferromagnets from neighboring layers. However, recent work has shown that when the spin current is generated in the bulk of a ferromagnet with appreciable spin-orbit coupling, both longitudinal and transverse spins can persist [27-29]. This indicates that spin dephasing is not relevant for all forms of spin current generation.

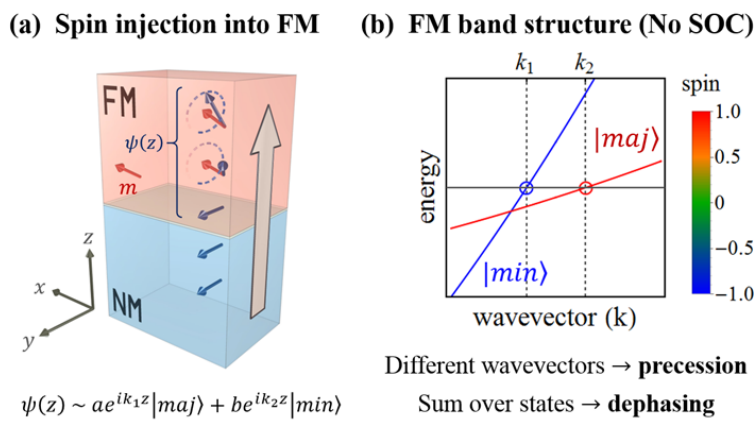

Figure 1. (a) Depiction of spin injection from a nonmagnetic layer into a ferromagnetic layer. When an electron with spin misaligned with the magnetization (**m**) enters the ferromagnetic layer, it must occupy a superposition of majority and minority states to preserve its transverse spin orientation. (b) Majority and minority bands of a model ferromagnet without spin-orbit coupling. Due to exchange splitting, majority and minority eigenstates (blue and red circles) have different Block wavevectors for the same energy, so the majority and minority eigenstates carry a phase difference given by $(k_1 - k_2)z$. For the superimposed state, this phase difference causes spin precession about the magnetization direction. Summed over all carriers, this spin precession is incoherent, leading to dephasing (loss of spin component transverse to magnetization).

To illustrate the conditions under which spin dephasing does or does not apply, we first explain its mechanism. Consider first the limit of vanishing spin-orbit coupling. As an example, if electrons with spin perpendicular to the magnetization are incident on a ferromagnet, the transmitted state is a superposition of majority and minority eigenstates. An example of this superposition of eigenstates is shown in Fig. 1(a). Due to exchange splitting in the ferromagnet's electronic structure (Fig 1(b)), majority and minority eigenstates with the same energy have different wavevectors (i.e. crystal momenta). Therefore, the transmitted electron is described by a superposition of majority and minority states with a phase difference that changes with position. This phase difference determines the electron's transverse spin direction, so as the phase difference varies in space the spin oscillates about the magnetization. Spin dephasing then occurs because precession amongst the injected spins is incoherent, so the net transverse spin density of all carriers rapidly vanishes. As a result, the net spin density points along the magnetization direction within a few atomic layers of the interface.

The addition of spin-orbit coupling changes this picture in multiple ways. Spin-orbit coupling in solids is predominantly an atomic-like, on-site potential of the form **L** · **s,** where **L** is the orbital angular momentum and **s** is the spin. The orbital character depends on Bloch wave vector **k**, so that spin-orbit coupling can be viewed as a **k**-dependent effective magnetic field. In general, the spin-orbit-derived magnetic field is not aligned with the magnetization. Any eigenstate's spin expectation value is aligned to the total effective magnetic field (magnetic exchange field + effective spin-orbit field), and is therefore not generally aligned with the magnetization [27]. In equilibrium and for the magnetization oriented along a high symmetry direction of the crystal (e.g. along an easy-axis), the *net* spin density of the occupied states is aligned with the magnetization and there is no torque on the magnetization. For a magnetization that is misaligned from an easy-axis, a torque induces precession about the easy axis, manifesting magnetocrystalline anisotropy. The magnetic force theorem relates this torque to the change in the spin-orbit energy of occupied states when the magnetization deviates from its easy-axis orientation [30]. The anisotropy torque can also be computed from the small transverse net spin density calculated non-self-consistently for a magnetization oriented away from an easy-axis [31].

The misalignment between electron spin and magnetization has important implications for electrically generated spin currents. In the picture of dephasing presented above, incoherent precession leads to the destruction of carrier spins transverse to the magnetization. In other words, the spin direction of injected electrons converges to the spin direction of bulk electrons within a few atomic layers. However, for ferromagnets with spin-orbit coupling, the spin directions of bulk electrons are misaligned with the magnetization, so theoretically transverse spin directions can survive dephasing. This phenomenon is seen in Ref. [11], where the authors present theoretical calculations showing that ferromagnets generate a substantial spin current flowing perpendicularly to the electric field and polarized transversely to the magnetization. This spin current results in part from the misalignment between electron spin and magnetization caused by spin-orbit coupling. Those calculations omit the extrinsic mechanisms associated with the spin Hall effect (skew scattering and side jump), so these spin currents arise only from the nonequilibrium occupation of the misaligned spin eigenstates.

However, for intrinsic mechanisms, where the electric field perturbs carrier wavefunctions rather than changing carrier occupation (see Fig. 2), dephasing is not relevant. In the intrinsic mechanism, the applied electric field couples an occupied and an unoccupied eigenstate with *different* energies at the same wavevector. Thus, the spin direction does not exhibit spatial oscillations as in the dephasing scenario described earlier, because the perturbed state has only a single wavevector. For this reason, intrinsically generated spin currents are not subject to dephasing, and a transverse component of the spin direction can exist in the bulk.

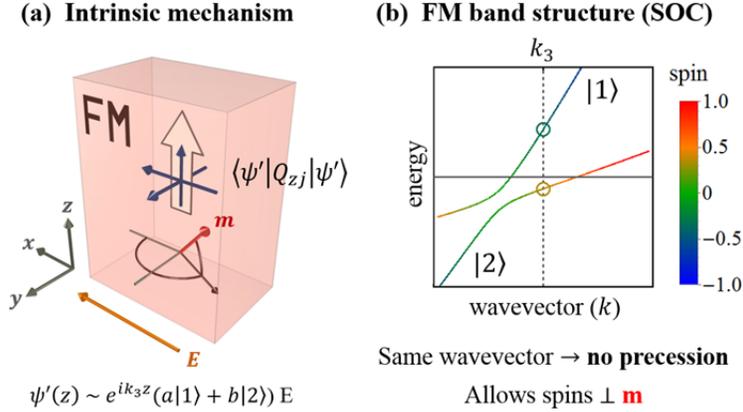

Figure 2. (a) Depiction of electrical spin current generation in a ferromagnet with magnetization (**m**) via the intrinsic mechanism. The applied electric field (**E**) generates a spin current, shown here for all three spin polarizations (blue arrows) with flow direction along z (block arrow). The perturbation to the electron wavefunction $\psi'$ from the electric field is given by a linear combination of the unperturbed eigenstates at the same wavevector. The resulting spin current can be computed (for vanishing disorder broadening) using the spin current operator $Q_{ij} \propto v_i \otimes \sigma_j$, where $v_i$ is the velocity operator, $\sigma_j$ are the Pauli matrices, and $i, j \in [x, y, z]$. (b) Band structure of a model ferromagnet with spin-orbit coupling. Note that the spin expectation value of the bands (taken from the colored bar legend) refers to the spin component along the magnetization. With spin-orbit coupling, the spin of eigenstates is not parallel/antiparallel to the magnetization in general due to the effective spin-orbit field. When the electric field perturbs the wavefunctions of bulk electrons, occupied and unoccupied states (yellow and green circles) are coupled at the same wavevector, so that dephasing is not relevant to transport. Extrinsic mechanisms (based on impurity scattering) can also lead to transverse spins (discussed below), but dephasing will play a role.

In contrast, extrinsic mechanisms are driven by impurity scattering. For instance, skew scattering, side jump, and the planar Hall effect all generate charge currents flowing perpendicularly to the electric field in ferromagnets. Since charge currents are spin-polarized in ferromagnets, it is possible that these mechanisms also create spin-polarized currents [32]. Spin swapping [10, 37, 41] describes the rotation of spins about the spin-orbit field of impurities during scattering events, which could generate transverse spin polarizations in ferromagnets. At ferromagnet/nonmagnet interfaces, spin filtering and spin precession due to the interfacial spin-orbit field results in spin current generation as well, referred to as interface-generated spin currents [11]. For these extrinsic mechanisms, more work is required to determine the role of dephasing.

These examples of spin current generation in ferromagnets provide only a limited view of the wide array of possible mechanisms. The rest of this article aims to cover the known examples of charge-spin conversion in ferromagnets driven by spin-orbit coupling. In section 2, we discuss what spin currents are allowed by symmetry in a ferromagnetic material. Based on this analysis, we group spin currents in ferromagnets based on their spin direction being longitudinal or transverse to the magnetization direction. We then review recent theoretical (sections 3-4) and experimental (sections 5-8) advances and classify them into these categories. Finally, in section 9, we discuss unanswered questions and new directions within this field.

2. **Symmetry-based argument**

The Curie principle [33] allows for or prohibits possible system responses based on symmetries. It requires that the symmetry of an effect (e.g. electrical spin current generation) must coincide with the symmetry of the cause (e.g. the symmetries of the material and applied electric field). If the effect breaks a symmetry that the cause preserves, then the effect cannot exist. If the cause breaks a symmetry then any effect that breaks the same symmetry is permissible.

We first consider the spin Hall effect in nonmagnetic materials to demonstrate the symmetry argument based on the Curie principle. A crystal with cubic symmetry, or a polycrystalline/amorphous material possesses inversion symmetry, two-fold (i.e. 180°) rotational symmetries about $x$-, $y$- and $z$-axes, and mirror symmetries about $xy$, $xz$, and $yz$ -planes. When an electric field is applied in the $x$-direction, it breaks the inversion symmetry, the two-fold rotational symmetries about $y$- and $z$-axis, and the mirror symmetry about the $yz$-plane. The system still possesses two-fold rotational symmetry about the $x$-axis ($C_2^x$) and the mirror symmetries about the $xz$ and $xy$ planes ($\sigma_{xz}$ and $\sigma_{xy}$). The Curie principle states that an effect due to an applied electric field in the $x$-direction must satisfy these three remaining symmetries.

Based on the symmetry argument above, an electric field along the $x$-direction ($j_x$) can only generate transversely flowing charge currents ($j_y$ or $j_z$) and spin currents ($\mathbf{Q}_y$ or $\mathbf{Q}_z$) as outlined in Figs. 1 and 2. Note that the spin current is a tensor $Q_{\alpha\beta}$ with two spatial indices: the subscript $\alpha$ specifies the flow direction and superscript $\beta$ specifies the spin direction. Here we use the notation that the spin current flowing in the $\alpha$-direction is a vector in spin space denoted by $\mathbf{Q}_\alpha$.

For example, in Fig. 3(a), the generated charge current violates the $\sigma_{xy}$ and $C_2^x$ symmetries, hence is not allowed in this system. Similarly, $Q_{zx}$ in Fig. 3(c) and $Q_{zz}$ in Fig. 3(e) are also forbidden by symmetry. $Q_{zy}$ in Fig. 3(d) is allowed by all symmetries, thus can exist. This configuration corresponds to the spin Hall effect, since the spin current $Q_{zy}$ represents flow along $z$ and spin direction along $y$, both of which are orthogonal to each other and the generating electric field along $x$. Note that this symmetry analysis does not specify either the strength of the effect or the microscopic mechanism. Nonmagnetic materials that break additional symmetries due to their crystalline structure have been shown to generate spin currents with spin direction different than the conventional spin Hall effect [34, 35].

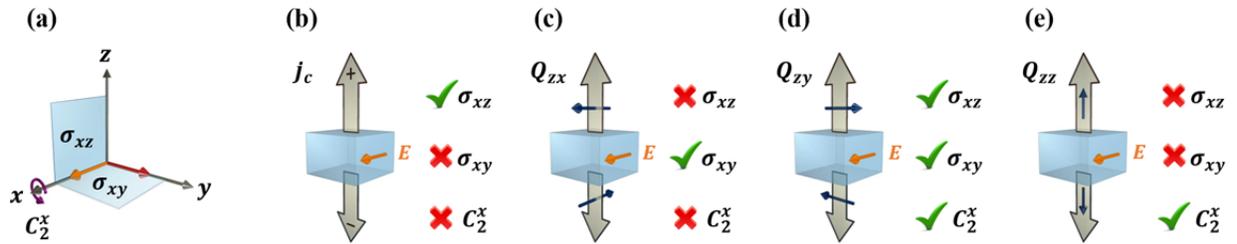

Figure 3. Depiction of the allowed electrically-generated spin currents in a nonmagnetic material. Panel (a) illustrates the two mirror plane ($\sigma_{xy}$ and $\sigma_{xz}$) and two-fold rotation ($C_2^x$) symmetries of a nonmagnetic material with an applied electric field along $x$. In panels (b)-(e), an effect is illustrated on the left and the symmetries it possesses or violates are shown on the right. Light blue rectangles represent the nonmagnetic films, orange arrows represent the applied electric field $\mathbf{E}$, grey block arrows represent the flow of charge in (a) and spin in (b)-(e). Dark blue arrows represent the spin direction of the spin currents. Panel (b) shows that an applied electric field along $x$ cannot generation additional charge flow in the $z$-direction. Panels (c)-(e) show that the only allowed spin current flowing in the $z$-direction has spin direction in the $y$-direction. This effect is phenomenologically identical to the spin Hall effect.

In the case of a ferromagnetic material, the magnetization breaks additional mirror symmetries about the planes which contain the magnetization and rotational symmetries about the axes perpendicular to the magnetization. Therefore, magnetic materials impose fewer constraints on the orientation of electrically driven spin currents allowed. For example, a magnetic material with magnetization along $y$ and electric field along $x$ has broken the $\sigma_{xy}$ and $C_2^x$ symmetries, leaving only the $\sigma_{xz}$ symmetry (Fig. 4(b)). Lifting the restrictions from $\sigma_{xy}$ and $C_2^x$ symmetries allows the generation of charge current $\mathbf{j}_c$, which corresponds to the well-known anomalous Hall effect [36]. By performing mirror reflection operation about the $xy$-plane (not shown), one finds that the anomalous Hall effect should be odd with magnetization, i.e. the induced current $\mathbf{j}_c$ switches direction as magnetization reverses. The generation of $Q_{zy}$ was allowed originally in nonmagnetic materials with more constraints, and therefore is still allowed here in magnetic materials. It can also be verified that $Q_{zy}$ is even in the magnetization by performing a mirror reflection operation about the $xy$-plane. The spin current $Q_{zy}$ corresponds to the spin Hall effect in ferromagnets [37], where the spin direction is longitudinal to the magnetization. The generation of spin currents $Q_{zx}$ and $Q_{zz}$ are still forbidden in this configuration.

For the case where the magnetization is aligned in the $z$-direction, as shown in Fig. 4(c), the only constraint is the $\sigma_{xy}$ symmetry. Therefore, both the generation of $Q_{zx}$ and $Q_{zy}$ are allowed, with the former being odd with respect to the magnetization and the latter being even with respect to the magnetization. The generation of $Q_{zy}$ follows the conventional spin Hall symmetry, but the spin direction $y$ is now transverse to the magnetization. The spin direction of $Q_{zx}$ can be described as $(\mathbf{z} \times \mathbf{E}) \times \mathbf{m}$, which is also transverse to the magnetization. Similarly, for the case where the magnetization is aligned in the $x$-direction, as shown in Fig. 4(a), the only constraint is the $C_2^x$ symmetry. Both the generation of $Q_{zy}$ and $Q_{zz}$ are allowed, with the former being even with respect to the magnetization and the latter being odd with respect to the magnetization. In this configuration, the spin direction of $Q_{zz}$ can also be described as $\mathbf{z} = (\mathbf{z} \times \mathbf{E}) \times \mathbf{m}$.

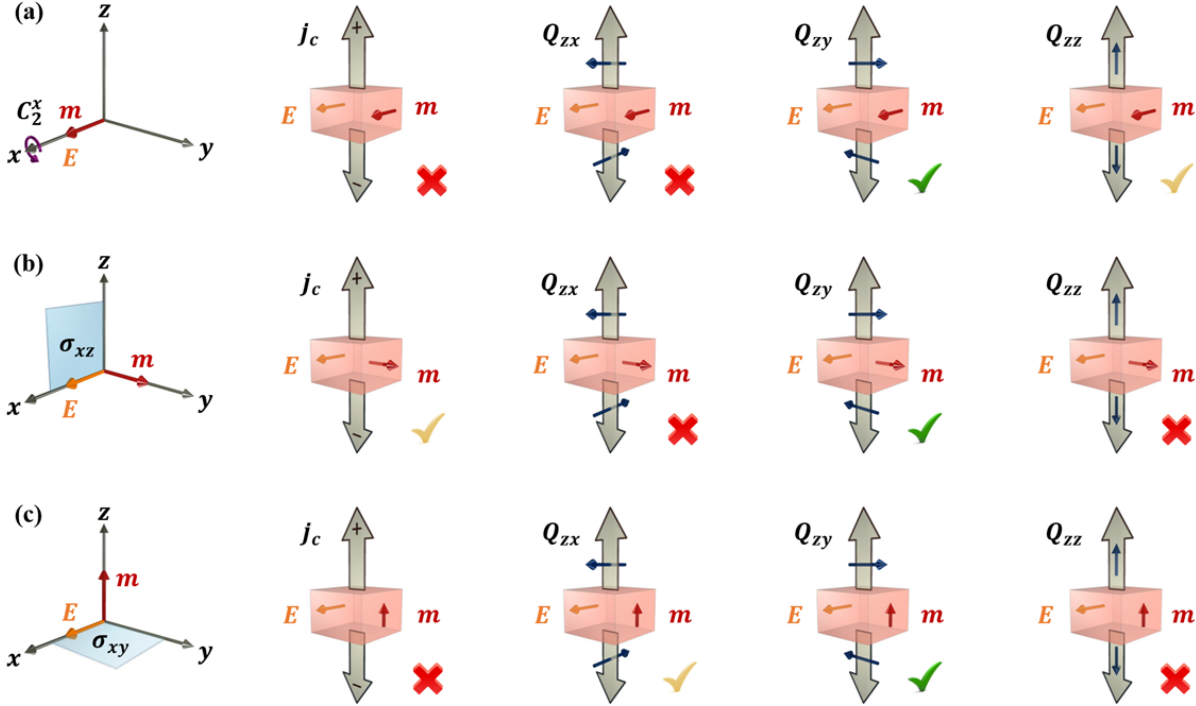

Figure 4. Symmetry analysis of the possible electrically-generated spin currents in a ferromagnetic material with three different magnetization orientations, each shown in panels (a)-(c). In each panel, the symmetry of the system is shown on the left. To the right, four hypothetical charge/spin currents are presented. Red crosses mean the spin current is disallowed by symmetry, green check marks mean the spin current is allowed by symmetry in both nonmagnetic and magnetic materials, and yellow check marks mean the spin current is allowed by symmetry in magnetic materials but not in nonmagnetic materials. Red arrows give the magnetization direction and the light blue rectangles represent the ferromagnetic thin films. The rest of the arrows follow the same conventions as in Fig. 3.

Since the transport behavior of longitudinal and transverse spins in a ferromagnet is different, it is useful to separate electrical spin current generation in ferromagnets into these two categories. Because it is impossible to distinguish detailed microscopic mechanisms based on the symmetry analysis alone, we use labels like the 'spin Hall effect' or 'spin anomalous Hall effect' to identify certain orientations of spin flow and spin direction but not to denote any particular microscopic mechanism. Because the existence of spin currents can be established by symmetry arguments, the effects discussed in this article should be universal for all ferromagnetic and ferrimagnetic conductors.

Based on the symmetry analysis provided above, we write a general expression for the spin current generated in ferromagnets. We assume the electric field **E** is applied in the film plane while the magnetization **m** is in an arbitrary direction (see Fig. 2). The generated spin current flowing in the $z$-direction can be expressed as

$$\mathbf{Q}_z = \sigma_{||}[\mathbf{m} \cdot (\hat{\mathbf{z}} \times \mathbf{E})]\mathbf{m} + \sigma_\perp \mathbf{m} \times [(\hat{\mathbf{z}} \times \mathbf{E}) \times \mathbf{m}] + \sigma_\perp^R \mathbf{m} \times (\hat{\mathbf{z}} \times \mathbf{E}), \tag{1}$$

where $\sigma_{||}$ is the conductivity associated with longitudinally-polarized spin currents and both $\sigma_\perp$ and $\sigma_\perp^R$ are conductivities associated with the two components of the transversely polarized spin currents. Eq. (1) satisfies all the symmetry relations specified in Fig. 4 when **m** is along $x$, $y$, and $z$ and allows for an arbitrary spin direction for all other directions of **m**, which is also consistent with the symmetry analysis. Note that the conductivity parameters are magnetization-dependent in general, but in certain cases could be well approximated as magnetization-independent. By dividing these three conductivities by the electric conductivity, we can also obtain the spin-Hall-angle-like angles $\theta_{||}$, $\theta_\perp$, and $\theta_\perp^R$.

Due to Onsager's principle, a spin current $\mathbf{Q}_z$ flowing in the $z$-direction into a ferromagnetic material with magnetization **m** can also generate electric currents:

$$\mathbf{j}_e = \theta_{||}(\mathbf{m} \cdot \mathbf{Q}_z)\mathbf{m} \times \hat{\mathbf{z}} + \theta_\perp (\mathbf{m} \times (\mathbf{Q}_z \times \mathbf{m})) \times \hat{\mathbf{z}} + \theta_\perp^R (\mathbf{m} \times \mathbf{Q}_z) \times \hat{\mathbf{z}}. \tag{2}$$

Below we summarize recent theoretical and experimental advances on electrically generated spin currents in ferromagnetic materials and the inverse effects.

## 3. Theory of electrical spin current generation in the bulk of ferromagnets

In this section we discuss some of the mechanisms responsible for spin current generation in bulk, centrosymmetric ferromagnets. We first consider the spin-polarized versions of the well-known Hall effects in ferromagnets (e.g. the anomalous Hall effect and the planar Hall effect). These spin-polarized currents have spin direction aligned with the magnetization. We then discuss how spin currents with spin direction *transverse* to the magnetization can form via extrinsic and intrinsic mechanisms. In general, both intrinsic and extrinsic mechanisms can create spin currents with spin directions longitudinal *and* transverse to the magnetization.

The anomalous Hall effect, discovered in 1880 by Edwin Hall, precedes many of the spin-orbit effects being studied today. The anomalous Hall effect describes a large magnetization-dependent Hall effect in a ferromagnetic conductor. After a century-long research effort, there is consensus that the microscopic mechanisms for the anomalous Hall effect include both an intrinsic mechanism (from the Berry curvature of the electronic structure) and extrinsic mechanisms (skew and side jump scattering off impurities) [38]. The relative contributions of these mechanisms to the anomalous Hall conductivity has been extensively studied and reviewed [30]. There is consensus that extrinsic mechanisms dominate in the limit of high conductivity (e.g. when the $\ell a \gg 1$, where $\ell$ is the mean free path and $a$ is the lattice constant), intrinsic mechanisms are important (and perhaps dominant) in the moderate conductivity range ($\ell a \approx 1$), while the mechanisms for the low conductivity range ($\ell a \ll 1$) is still not fully understood.

We first describe the anomalous Hall response of a ferromagnet due to impurity scattering (extrinsic mechanism), which can be understood in terms of a semiclassical theory. Zhang [37] used a drift-diffusion model to show that the anomalous Hall effect can be viewed as a special case of the spin Hall effect, as sketched in Fig. 5(a). In a ferromagnet, the band structure of majority and minority spins are significantly different, leading to different spin Hall conductivities for these two spin species, $\sigma_\uparrow^{SH}$, and $\sigma_\downarrow^{SH}$, where the arrows in the subscripts represent majority and minority spins respectively. Note that $\sigma_\uparrow^{SH}$ corresponds to flow direction along $\mathbf{E} \times \mathbf{z}$, and $\sigma_\downarrow^{SH}$ corresponds to flow direction along $-\mathbf{E} \times \mathbf{z}$, where $\mathbf{E}$ is the applied electric field and $\mathbf{z}$ is the quantization axis for the ferromagnet. Therefore, the anomalous Hall current density can be expressed as

$$j_e^{AH} = \left(\sigma_\uparrow^{SH} - \sigma_\downarrow^{SH}\right)E. \tag{3}$$

Besides the electric current density, the anomalous Hall effect also leads to a spin current density, which can be expressed as

$$Q_s^{AH} = \left(\sigma_\uparrow^{SH} + \sigma_\downarrow^{SH}\right)E. \tag{4}$$

Note that we have written the spin current density in units of charge current density for ease of comparison with charge-based effects (multiplying by $\hbar/2e$ converts Eq. (4) to spin current density). This theoretical model shows that a ferromagnet can be used to generate spin current flowing transverse to the electric field, just like the spin Hall effect in nonmagnetic materials. An important assumption of this model is that only spin states collinear to the magnetization are considered. Therefore, in this model the spin direction must be parallel to the magnetization.

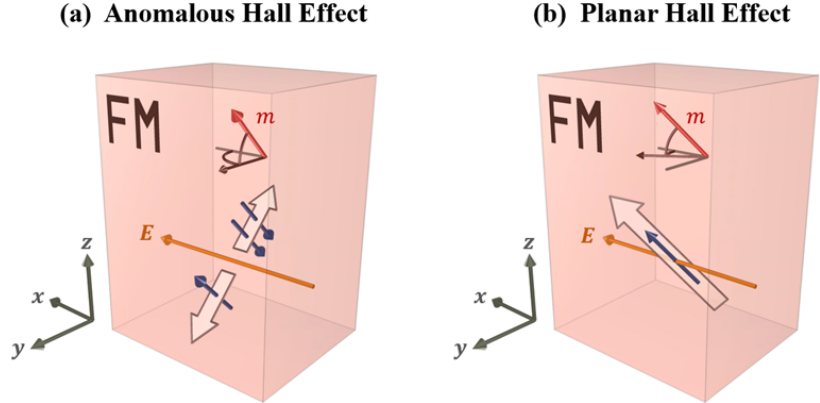

Figure 5. Illustrations of the (a) anomalous Hall effect and the (b) planar Hall effect in a ferromagnet under the assumption that all spins are colinear to the magnetization (**m**). Arrows follow the same convention as Figs. 1 and 2. Both effects can generate spin-polarized currents (Taniguchi *et al.* [32]). (a) For the anomalous Hall effect (as with the spin Hall effect), the electric field, spin flow, and spin direction are mutually orthogonal. Thus, for an electric field along *x*, the spin-polarized current flows along *z* if the magnetization (and thus spin direction) is along *y*. However, a spin current with both flow and spin direction along *z* can occur if the magnetization is tilted away from the *y*-axis but still carries a *y*-component. (b) When the magnetization is tilted away from the electric field but not perpendicular to it, a charge current forms that flows along the magnetization direction. This phenomenon is the mechanism behind the planar Hall effect. Since this charge current is spin-polarized, the planar Hall effect also gives rise to a spin current.

Based on similar reasoning, Taniguchi *et al.* [32] theoretically predicted that the magnetization can be used as an additional knob to control the spin direction of the spin current. This work considered extrinsic mechanisms, so that states with perpendicular spin components are superpositions of eigenstates with different Bloch wave vector, as described in Fig. 1. Due to strong dephasing, Taniguchi *et al*. argued the spin current generated by the anomalous Hall effect should always be polarized parallel to the magnetization. Therefore, when the magnetization has an out-of-plane component, as shown in Fig. 5(a), the spin-polarized anomalous Hall current should also contain an out-of-plane spin component. The magnitude of the spin current in this model is proportional to $m_y m_z$, where $m_y$ and $m_z$ are unitless magnetization components in the *y*- and *z*- directions. The component $m_y$ captures the magnetization dependence of the anomalous Hall effect and the component $m_z$ captures the out-of-plane component of the spin polarization. Spin currents with out-of-plane spin direction are potentially useful for magnetic memory applications since they could switch out-of-plane magnetizations, which are usually preferred for energy-efficient switching, high packing density, and scalability. Besides the anomalous Hall effect, Taniguchi *et al.* [32] also showed that the planar Hall effect – which arises from the anisotropic magnetoresistance – generates a spin current. The planar Hall current flows perpendicularly to the electric field but vanishes when the magnetization is perpendicular or parallel to the electric field. As illustrated in Fig. 5(b), the planar Hall current must be spin-polarized, and the magnitude of this spin current is expected to be proportional to $m_x m_y m_z$ for flow along *y* and spin polarization along *z*.

The spin currents in ferromagnets discussed so far originate from charge-based Hall effects under the assumption that the Hall currents are spin polarized along the magnetization. The symmetry arguments presented in section 2 show that this assumption is too restrictive. In what follows, we discuss some of the microscopic mechanisms studied so far that expand upon the picture presented above. These microscopic mechanisms reveal that spin planar Hall currents can have spin directions transverse to the magnetization, spin anomalous Hall currents must include (but are not limited to) an additional spin Hall component, and spin swapping could generate novel phenomena not captured by spin-dependent generalizations of previously known effects.

Amin *et al.* [11] used ab-initio based tight-binding models to show that single ferromagnetic layers generate spin currents flowing perpendicularly to the electric field with components of spin direction along $\mathbf{m} \times (\mathbf{f} \times \mathbf{E})$ and $\mathbf{m}$, where $\mathbf{f}$ gives the spin flow direction. The latter can be identified as the spin-polarized planar Hall effect described in Ref. [32]. However, a spin current with spin direction $\mathbf{m} \times (\mathbf{f} \times \mathbf{E})$ is allowed by symmetry as well, as can be seen by setting $\mathbf{f} = \mathbf{z}$ in Eq. 1. In Ref. [11], the occurrence of this transversely polarized spin current is traced back to the misorientation of the eigenstate spin with the magnetization due to the effective spin-orbit field. Note that the calculations in Ref. [11] do not yield all possible spin currents described in Eq. 1 because the effect of the electric field is limited to a perturbation in the occupation of carriers. The impurity potentials associated with skew scattering and side jump and the perturbation to electronic wavefunctions are ignored, eliminating the extrinsic and intrinsic mechanisms that give rise to the spin Hall effect and anomalous Hall effect. Nevertheless, the transport calculations in Ref. [11] reveal that both bulk and interfacial spin current generation can be significant. Surprisingly, the spin planar Hall effect with spin direction along $\mathbf{m} \times (\mathbf{f} \times \mathbf{E})$ is computed to be about 3000 $\Omega^{-1}\text{cm}^{-1}$ in Co, three times larger than its counterpart with spin direction along $\mathbf{m}$.

Next, we consider how the intrinsic mechanism underlying the anomalous Hall effect leads to spin current generation that qualitatively differs from the mechanisms we've described so far. The intrinsic contribution to the anomalous Hall conductivity can be expressed in terms of the Berry curvature of the electronic states occupied in equilibrium. Equivalently, the anomalous Hall effect can be viewed as the electric-field induced perturbation of the electron wavefunctions, which leads to the formation of a transverse charge current. The perturbed wavefunction can be expressed as a linear combination of Bloch states with the same Bloch wavevector **k**, as outlined in Fig. 1. This is a consequence of the long wavelength limit of the DC electric field perturbation (e.g. the limit $q \to 0$, where $q$ is the wave vector of the electric field): momentum conservation imposes interband coupling between states with equal **k**. The strongest contribution to the anomalous Hall effect arises from interband coupling between occupied and unoccupied states that are nearby in energy. If this pair of states have opposite spin character (or more precisely, are strongly modified by spin-orbit coupling), then interband coupling results in a spin current flowing transverse to the electric field, with spin direction transverse to the magnetization (see Fig. 1). This state is not subject to dephasing because it carries a single Bloch wavevector, rather than being a superposition of states with differing Bloch wavevectors.

The substantial intrinsic contribution to the anomalous Hall effect in transition metal ferromagnets suggests that intrinsic contributions to spin currents generated in ferromagnets should also be important. Recent first-principles calculations of the magnetization-dependent intrinsic spin current conductivity in Co, Fe, and Ni confirm that this is indeed the case [28, 29]. Ref. [28] showed that the intrinsic contribution is well-described by Eq. (1) for cubic crystals Fe and Ni, where $\sigma_\parallel$ and $\sigma_\perp$ are well approximated as magnetization-independent parameters. The magnitude of the computed spin Hall conductivity components are given by $\sigma_\parallel = 100$, $\sigma_\perp = 519$ for Fe, and $\sigma_\parallel = 960$, $\sigma_\perp = 1688$ for Ni (all values given in units $\hbar/2e$ $[\Omega^{-1} \cdot cm^{-1}]$. Hcp Co is not as well-described by Eq. (1) because of its substantial crystal anisotropy. Note that the authors find that the intrinsic contribution to $\sigma_\perp^R$ vanishes for all materials. The $\sigma_\perp^R$ term vanishes because it must be odd under time-reversal, which can be seen through inspection of Eq. (1) and noting that the spin current and electric field are even under time-reversal. If a force and response transform differently under time-reversal, the physical mechanism requires dissipation [38]. The intrinsic mechanism is dissipationless and therefore returns a vanishing conductivity $\sigma_\perp^R$.

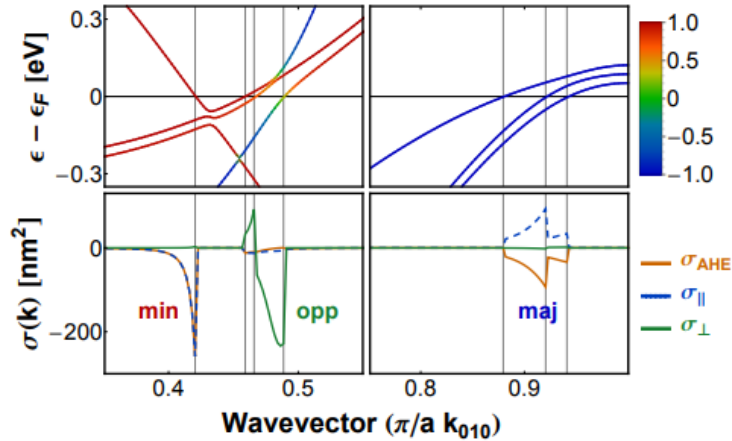

Figure 6 (Color online) Band structure near the Fermi energy (top) and k-dependent intrinsic conductivities (bottom) for BCC Fe, where $\hat{m} = (\hat{y} + \hat{z})/\sqrt{2}$. Band color gives value of $\mathbf{s} \cdot \hat{m}$, where $\mathbf{s}$ is the spin and blue (red) bands corresponding to majority (minority) carriers. Avoided crossings between like (opposite) spin bands contribute strongest to $\sigma_\parallel$ ($\sigma_\perp$), which describes the spin current with spin direction parallel (perpendicular) to $\hat{m}$. Images adapted from Ref. [28].

The top panels of Fig. 6 show the band structure of Fe near the Fermi level while the bottom panels show plots of the relevant charge and spin conductivity parameters. The pair of opposite-spin bands shown in Fig. 6 results in a peak in $\sigma_\perp$ (labelled "opp" in the figure), while like-spin bands result in peaks in $\sigma_\parallel$ and $\sigma_{AHE}$ (labelled "min" and "maj" in the figure). In general, $\sigma_\perp$ results entirely from interband coupling between spin-opposite bands, $\sigma_{AHE}$ results from coupling between spin-like bands, and $\sigma_\parallel$ has contributions from coupling between both spin-opposite and spin-like pairs. The dependence of these conductivities on the spin character of the coupled bands shows that, unlike the extrinsic case described by Zhang, there is not a simple relation between the intrinsic anomalous Hall conductivity and longitudinal spin Hall conductivity as described by Eqs. (3) and (4). This can be understood by noting that spin-opposite band pair contributions cannot be associated with a majority or minority conductivity.

So far, we have discussed how charge-based Hall effects generalize to spin-dependent effects in ferromagnets. However, there are other means to *indirectly* generate spin currents via an electric field in nonmagnetic and ferromagnetic materials. Lifshits and Dyakonov predicted that when spin currents are injected into nonmagnetic materials, impurity scattering creates a new spin current flowing in a different direction [10]. If the injected, or primary, spin current has different flow and spin directions, the new spin current will have those flow and spin directions swapped. For this reason, the effect is known as "spin swapping." For instance, if the injected spin current flows along *x* and has a spin direction along *y*, the spin swapping current flows along *y* with spin direction along *x*. If the injected spin current has the same flow and spin direction (say both along *x*), this leads to spin swapping currents with flow/spin direction along *y* and flow/spin direction along *z*. The effect predicted by Lifshits and Dyakonov results from impurity scattering; an intrinsic analogue was later introduced by Sadjina *et al*. [39]. Spin-orbit scattering at interfaces can also lead to the spin swapping effect [40-42].

The original proposal of spin swapping by Lifshits and Dyakonov focused on nonmagnetic materials. To introduce the primary spin current, the authors suggested running a charge current through a separate ferromagnetic layer and allowing the resulting spin-polarized current to flow through a spacer material into a nonmagnetic layer. A natural alternative is to consider a single ferromagnetic layer, in which spin-dependent scattering generates a spin polarized current and then spin-orbit scattering creates a spin swapping current, all within the same material. However, the spin swapping current, which has spin direction orthogonal to the magnetization, must survive dephasing to be measurable.

Ortiz Pauyac, Chshiev, *et al.* [43] used a quantum kinetic approach to show that transversely polarized spin currents can be generated within ferromagnets via impurity scattering. The microscopic mechanisms include side jump, skew scattering, spin swapping, spin relaxation, and Larmor precession. As illustrated in Fig. 7, an electric field is applied in the *x*-direction in a magnetic film with magnetization in the *y*-direction. Electrons flowing in the *x*-direction are spin polarized in the *y*-direction by the magnetization. The spin swapping effect gives rise to a spin current flowing in the *y*-direction with *x*-spin direction. This spin current alone could deposit an *x*-polarized spin accumulation at the edges (Fig. 7(b)), but spin precession about the magnetization generates an additional *z*-component. Skew scattering and side jump generate a spin current flowing in the *y*-direction with *z*-spin direction (Fig. 7(c)), which alone would deposit a *z*-polarized spin accumulation at the edges. Similarly, spin precession about the magnetization adds an *x*-component to this spin accumulation. The net effect, as shown in Fig. 7(d), is the generation of spin currents and spin accumulations with the conventional spin Hall orientation (*z*-component spin) and with a rotated spin orientation (*x*-component spin). Note that the competition between dephasing and spin-orbit scattering determines how far from the interface these spin accumulations survive. Strong dephasing could greatly reduce spin swapping effects deep within the bulk of the ferromagnet.

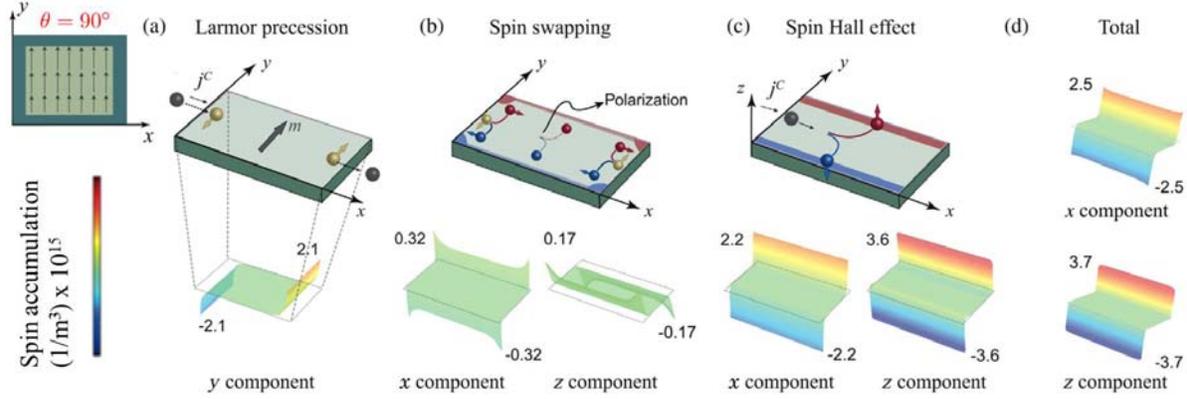

Figure 7. Current-induced spin accumulation in a ferromagnet due to (a) Larmor precession, (b) spin swapping and (c) the spin Hall effect (only including skew scattering and side jump). Images adapted from Ref. [43]. In all cases, the charge current is along $x$ and the magnetization is along $y$. Spin swapping, skew scattering, and side jump all generate spin accumulations at interfaces along the $x$ and $z$ directions (transverse to the magnetization).

## 4. Theory of electrical spin current generation at ferromagnet/nonmagnet interfaces

While the bulk properties of ferromagnets enable electrical spin current generation, the broken symmetries at ferromagnet/nonmagnet interfaces allow for additional effects. The Rashba-Edelstein effect is an important example and refers to an electric field-induced spin accumulation at the interface. To theoretically investigate electrically-generated spin currents at interfaces that flow out-of-plane requires a three-dimensional treatment of the region near the interface. Such a three-dimensional treatment is typically not considered when studying the Rashba-Edelstein effect, though the role of interfacial spin-orbit coupling on three-dimensional spin transport has garnered increasing attention [40-42, 44-48]. In the following, we discuss how interfaces modify incident spin currents, and how charge- spin conversion occurs at interfaces through spin swapping and spin-orbit scattering (i.e. interface-generated spin currents).

One way in which ferromagnet/nonmagnet interfaces can modify spin currents generated in bulk layers is via the exchange interaction. For example, when a spin current traversing the nonmagnetic layer reaches the interface, the scattered spins will precess due to the exchange interaction at the interface. The transmitted spins will dephase due to the bulk exchange interaction in ferromagnets while the reflected spins will have rotated relative to the incident spins. The reflected spin current has a spin direction with the following components: $\mathbf{m} \times (\mathbf{m} \times \mathbf{s})$, $\mathbf{m} \times \mathbf{s}$, and $\mathbf{m}$, where $\mathbf{s}$ is the direction of the incident spins. In the following, we consider only the first two components, which are both transverse to the magnetization. The amplitude of the transverse spin reflection can be succinctly parameterized by a complex-valued interface conductance, called the spin mixing conductance [49, 50], where the real and imaginary parts describe the reflected spins along the $\mathbf{m} \times (\mathbf{m} \times \mathbf{s})$ and $\mathbf{m} \times \mathbf{s}$ directions respectively. More precisely, if the interface is located at $z = 0$, with the nonmagnet at $z < 0$ and the ferromagnet at $z > 0$, then the spin mixing conductance relates the spin accumulation at the interface but just within the nonmagnet ($z = 0^-$) to the total spin current (incident plus reflected) at the same location:

$$\mathbf{Q}_z = G_{mix}\boldsymbol{\mu}_s. \tag{5}$$

Here, we momentarily depart from previous conventions and allow all variables in Eq. 5 to be complex-valued. Eq. 5 only describes spin directions oriented transversely to the magnetization. The spin current $\mathbf{Q}_z$ flows out-of-plane ($z$-direction) and the real and imaginary components of $\mathbf{Q}_z$ are the two components of the spin direction transverse to the magnetization. The real and imaginary components of the spin accumulation $\boldsymbol{\mu_s}$ likewise describe the two components of spin accumulation transverse to the magnetization. The real part of $G_{mix}$ describes the component of spin current with spin direction along $\mathbf{m} \times (\mathbf{m} \times \mathbf{s})$ while the imaginary part of $G_{mix}$ describes the spin direction along $\mathbf{m} \times \mathbf{s}$. Note that in this model, the transverse spin accumulation and spin current in the ferromagnet are assumed to vanish due to dephasing, so only the spin accumulation and spin current in the nonmagnet are relevant for transport.

The spin mixing conductance describes the modification of spin currents incident to nonmagnet/ferromagnet interfaces but does not describe any direct coupling to an external electric field. Thus, the relevance of the spin mixing conductance here is limited to cases where an electric field generates a spin current and the spin mixing conductance modifies that spin current at an interface. A notable example occurs in heavy metal/ferromagnet bilayers driven by an in-plane electric field, where the spin Hall effect generates a spin current in the heavy metal flowing out-of-plane and the spin mixing conductance modifies that spin current near the interface. The incident spin current has spin direction $\mathbf{s} = \mathbf{z} \times \mathbf{E}$ and the reflected spin current has components of spin direction along $\mathbf{m} \times (\mathbf{m} \times \mathbf{s})$ and $\boldsymbol{m} \times \boldsymbol{s}$, where the strength of these components are mostly determined by the real and imaginary parts of the mixing conductance respectively.

In nonmagnet/ferromagnet bilayers under an in-plane electric field, spin swapping also results in spin currents with spin direction $\boldsymbol{s'} = \mathbf{m} \times \mathbf{s}$, as was outlined by Saidaoui and Manchon [51]. A simple way to understand the role of spin swapping follows by assuming the magnetization points along the *z*-direction. If the applied electric field is along the *x*-direction, then the spin-polarized current in the ferromagnet flows along *x* and has spin direction along *z*. If some of this current enters the nonmagnetic layer, the resulting spin swapping current has flow along *z* and spin direction along *x*, which can be written as $\boldsymbol{s'} = \mathbf{m} \times \mathbf{s}$ where $\mathbf{s} = \mathbf{z} \times \mathbf{E}$ as before. In general, electrons carry a spin polarization along $-\mathbf{m}$ in the ferromagnet, and one can show that those electrons which scatter into the nonmagnet experience a net spin-orbit field along $\boldsymbol{s}$ from impurities via spin swapping. This causes the electron spins along $-\mathbf{m}$ to precess about the effective spin-orbit field $\boldsymbol{s}$, yielding a new component of spin polarization along $\boldsymbol{s'} = \mathbf{m} \times \mathbf{s}$. This effect vanishes if the nonmagnet layer is greater than a mean free path, so spin currents generated in this manner cannot fully traverse nonmagnetic layers greater than a mean free path [51].

Both effects described above rely on the presence of spin-orbit coupling in the nonmagnetic layer and assume that spin-orbit coupling is negligible at the interfaces. However, spin currents can be generated via electric field through coherent spin-orbit scattering at interfaces. For nonmagnetic interfaces, Linder and Yokoyama [52] demonstrated that a charge current injected perpendicular to an interface (**z**) generates a spin current that flows in a direction $\boldsymbol{f}$ parallel to the interface plane with spin direction $\boldsymbol{f} \times \mathbf{z}$. This process is loosely analogous to the bulk extrinsic spin Hall effect, where the role of the impurity has been replaced by the interface.

Spin current generation at interfaces was later explored in nonmagnet/ferromagnet bilayers by Amin and Stiles [40, 41] under different assumptions than [52], in which the electric field is parallel to the interface plane and generates a spin current flowing perpendicular to the interface plane. In their work, these "interface-generated spin currents" were shown to exert spin torques on the ferromagnetic layer. Furthermore, the presence of the ferromagnet breaks additional symmetries as compared to nonmagnetic interfaces, enabling spin currents pointing in all directions.[2] Together with Zemen [11], the strength of interface-generated spin currents were calculated using tight-binding models fitted to ab-initio band structures for Co/Pt, Co/Cu, and Pt/Cu interfaces. Note that the Pt/Cu system is nonmagnetic, and symmetry dictates that the interface-generated spin current must have spin direction along $\mathbf{s} = \mathbf{z} \times \mathbf{E}$. For the nonmagnet/ferromagnet interfaces, like Co/Pt and Co/Cu, symmetry dictates that an arbitrary magnetization direction leads to a spin current with spin direction in three directions: $\mathbf{s} = \mathbf{z} \times \mathbf{E}$, $\mathbf{m} \times \mathbf{s}$, and $\mathbf{s} \times (\mathbf{m} \times \mathbf{s})$. Fig 8(a) depicts the allowed interface generated spin currents in these systems.

---

[2] These effects can be thought of loosely as inverse effects, but note that the true inverse effect of the mechanism predicted by Linder and Yokoyama would involve a spin current flowing along the interface generating a charge current flowing out-of-plane. Whereas Onsager reciprocity requires the direct and inverse effects to have the same

Interface-generated spin currents have been confirmed by Freimuth *et al.* [53] using more sophisticated density functional theory calculations. Because these spin currents arise from coherent spin-orbit scattering at interfaces, they are not restricted by the thickness of the nonmagnetic layer like the spin swapping effect above [51], and could in principle traverse nonmagnetic layers greater than a mean free path. However, interface-generated spin currents do scale with the conductivities of the bulk layers, unlike intrinsic effects which are independent of the impurity concentration [11].

Using toy models provides some intuition about the physical origin of interface-generated spin currents. Consider a simple model in which both the nonmagnet and ferromagnet are modeled as free electron gasses with identical, spherical, spin-independent Fermi surfaces. In the ferromagnetic layer, the imbalance of majority and minority carriers enters through a spin-dependent nonequilibrium occupation of carriers. The spin-dependent interfacial potential is given by

$$V(\mathbf{r}) \propto \delta(z)(u_0 \mathbf{I}_{2\times 2} + u_R \boldsymbol{\sigma} \cdot (\hat{\mathbf{k}} \times \hat{\mathbf{z}}))$$

where $u_0$ gives the spin-independent barrier, $u_R$ is the scaled Rashba parameter, $\boldsymbol{\sigma}$ is the vector of Pauli matrices, and $\hat{\mathbf{k}}$ is the incident wavevector. In this case, the general form of the interface-generated spin current simplifies considerably and can be described by two effects: spin-orbit filtering and spin-orbit precession. As shown in Figs. 8(b-c), electrons with wavevector $\mathbf{k}$ scattering off the interface will briefly interact with a Rashba spin-orbit field given by $\mathbf{u}(\mathbf{k}) = \mathbf{z} \times \mathbf{k}$. Spins that are aligned or antialigned with $\mathbf{u}(\mathbf{k})$ have different reflection and transmission probabilities; in this scenario $\mathbf{u}(\mathbf{k})$ behaves a k-dependent spin filter, which describes spin-orbit filtering. Spins that are misaligned with $\mathbf{u}(\mathbf{k})$ will additionally precess upon scattering, which describes spin-orbit precession. The combination of these two effects fully describes the spin current generated by spin-orbit scattering at the interface in this simple model, where spin-orbit filtering yields a spin direction $\mathbf{s}$ and spin-orbit precession yields a spin direction $\mathbf{m} \times \mathbf{s}$.

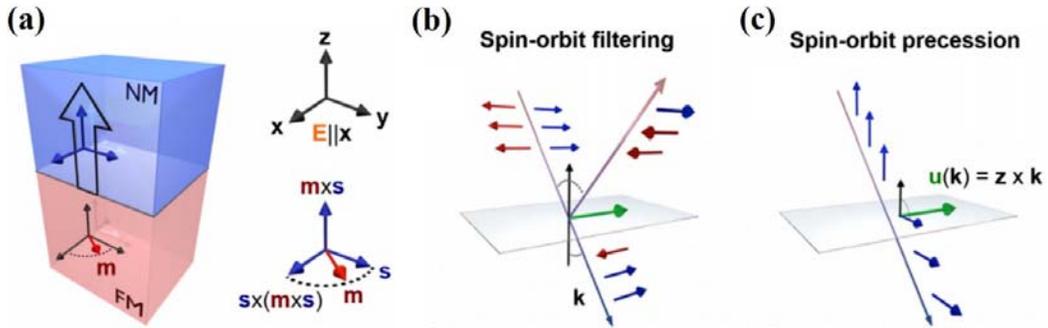

Figure 8 (a) Spin current generation near a ferromagnet/nonmagnet interface. (b-c) Illustration depicting spin-orbit filtering and spin-orbit precession. Here the red and blue arrows represent spin moments and the green arrows represent the interfacial spin-orbit field $\mathbf{u}(\mathbf{k}) = \mathbf{z} \times \mathbf{k}$. Images adapted from Ref. [11].

---

strength, the effect predicted by Linder and Yokoyama and the effect predicted by Amin and Stiles can have different values.

The transport calculations in Ref. [11] reveal that the conductivities describing interface-generated spin currents with spin direction along **s**, **m** × **s**, and **s** × (**m** × **s**) are significant, sometimes exceeding 1000 $\Omega^{-1}$cm$^{-1}$ at both Co/Pt and Co/Cu interfaces. This suggests that interfaces are important sources of spin current that could compete with spin current generation from bulk layers.

## 5. Explanation of Terminology in Reviewing Experimental Results

Before we review recent experimental observations of charge-spin conversion in ferromagnets, we explain our use of terminology. As discussed in the theory sections above, when a charge current passes through a ferromagnet, a spin current can be generated in the bulk of the ferromagnet as well as at the interface between the ferromagnet and a neighboring layer. It is challenging to experimentally distinguish between the bulk and interface-generated spin currents because both can have similar dependencies on the magnetization direction. Therefore, in the experiment sections below, we use the term "spin Hall effect" to refer to the generation of a spin current from either a ferromagnet or a ferromagnet/nonmagnet interface where the flow and spin directions follow the conventional spin Hall orientation. We emphasize that, in the interpretation of experiments, interfacial spin current generation could be mistaken as bulk spin current generation and vice versa. We will use longitudinal spin Hall effect, transverse spin Hall effect and transverse spin Hall effect with spin rotation to describe the different spin-charge conversions described by the three terms in Eq. (1), respectively. For the first term, $\sigma_{||}[\mathbf{m} \cdot (\hat{\mathbf{z}} \times \mathbf{E})]\mathbf{m}$, "longitudinal" means spin direction is along magnetization **m**. For the second term, $\sigma_{\perp}\mathbf{m} \times [(\hat{\mathbf{z}} \times \mathbf{E}) \times \mathbf{m}]$, "transverse" means the portion of $\hat{\mathbf{z}} \times \mathbf{E}$ that is perpendicular to **m**, where $\hat{\mathbf{z}}$ is the spin current flow direction. For "transverse with spin rotation" in the third term, $\sigma_{\perp}^R \mathbf{m} \times (\hat{\mathbf{z}} \times \mathbf{E})$, the spin direction is perpendicular to **m**, and perpendicular to $\hat{\mathbf{z}} \times \mathbf{E}$. The corresponding inverse effects will be called the inverse longitudinal spin Hall effect, inverse transverse spin Hall effect, and inverse transverse spin Hall effect with spin rotation, which correspond to the three terms in Eq. (2), respectively. We emphasize the use of "spin Hall effect" in reviewing the experimental results does not imply a particular microscopic mechanism.

## 6. Experimental Observation of Conversion between Charge Currents and Longitudinal Spins

Charge-spin conversion in a nonmagnetic material can be detected in a spin Hall effect configuration, in which an applied charge current generates a flow of spin current. The spin current may apply a spin torque on a neighboring ferromagnetic material [9] or generate a chemical potential difference near an interface with a ferromagnetic material [6]. The charge-spin conversion can also be detected in an inverse spin Hall effect setting, where a spin current is generated from a ferromagnetic material via the spin pumping effect [8] or spin Seebeck effect [54] and injected into the nonmagnetic material. The inverse spin Hall effect in the nonmagnetic material converts the spin current into an electric current. In either case, a ferromagnetic material is often needed for measuring the charge-spin conversion in a nonmagnetic material. Therefore, a spin valve structure is often used for measuring the charge-spin conversion in the ferromagnetic material, where one of the magnetic layers serves as a spin current generator and the other magnetic layer as a spin current detector.

Miao *et al.* [55] studied charge-spin conversion in Py by using a YIG/Py bilayer. As shown in Fig. 9(a). The spin Seebeck effect [54] from YIG injects a spin current into Py, which then generates in-plane electric fields, due to the anomalous Nernst effect from Py, $E_{\text{ANE}}$, and the inverse spin Hall effect from Py, $E_{\text{ISHE}}$. In this geometry, the injected spin polarization is parallel with Py magnetization, therefore, the relevant process is the inverse longitudinal spin Hall effect. Shown in Fig. 9(b), the voltage signal due to $E_{\text{ISHE}}$ from the YIG/Py sample is extrapolated by deducting the voltage signal measured in two control samples, YIG/MgO/Py and YIG(Ion bombarded)/Py. In both control samples, spin injection from YIG to Py is suppressed, hence the voltage signal shall only consist of the contribution from $E_{ANE}$ of Py. The net voltage signal due to $E_{ISHE}$ is compared to the spin Seebeck voltage signal measured in a YIG/Pt bilayer, from which an effective $\theta_{\parallel}$ of Py is found to be 38% of the spin Hall angle of Pt. Wu *et al.* [56] carried out a similar experiment using a structure YIG/Cu/Py/IrMn, where Cu is used to separate direct exchange coupling between Py and YIG and IrMn is an antiferromagnet that applies an exchange bias on the Py magnetization. In this way, the magnetization hysteresis of YIG can be distinguished from that of Py. The thermal voltage signal in the multilayer tracks the YIG magnetization hysteresis, and corresponds to the inverse spin Hall effect. The effective $\theta_{\parallel}$ extrapolated from Wu's experiment is 98% of the Pt spin Hall angle.

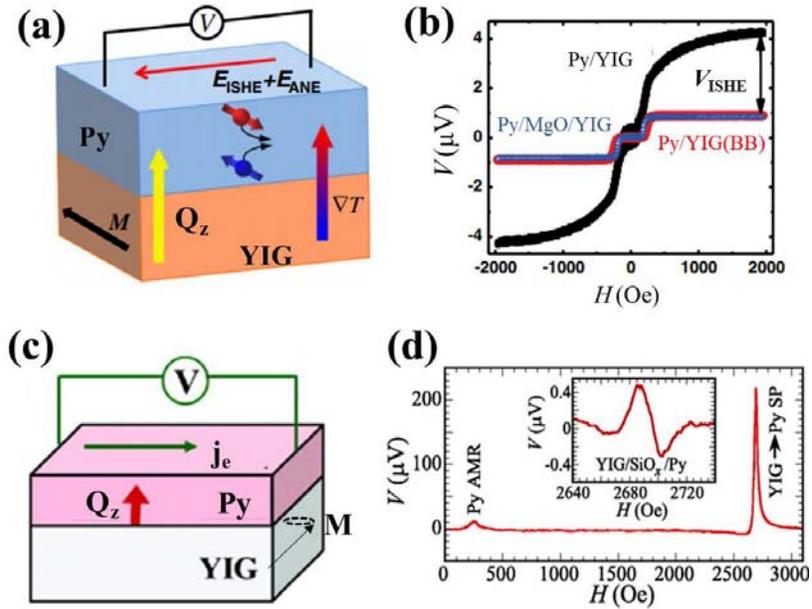

Figure 9 (a) Illustration for the spin Seebeck effect measurement. (b) Measured thermal voltage in Py/YIG, Py/MgO/YIG and Py/YIG (ion bombarded). The former contains both ANE signal of the Py and the longitudinal spin-charge conversion due to the spin current injected from YIG to Py. The latter two only contains the ANE signal of Py. (c) Illustration of the spin pumping measurement which utilizes longitudinal spin-charge conversion. (d) Measured voltage signal. The signal corresponding to the YIG resonance is attributed to the longitudinal spin-charge conversion due to the spin current pumped from YIG to Py. Images adapted from Refs. [55] and [56].

Wang *et al.* [57] studied charge-spin conversion using the DC spin pumping effect in a YIG/Cu/Py trilayer. As shown in Fig. 9(c-d), by applying microwave magnetic field to the sample, a large DC voltage signal corresponding to the magnetic resonance of YIG is measured in the metallic layer, which is attributed to the inverse spin Hall effect in Py. Since the spin polarization that generates the DC voltage is parallel with the Py magnetization, DC voltage is also due to the inverse longitudinal spin Hall effect. The effective $\theta_{\parallel}$ for Py is extrapolated to be 0.02. It is worth noting that spin pumping not only gives rise to a DC part of spin polarization, but can also generate an alternating spin polarization transverse to the Py magnetization, which may generate a voltage in the metallic layer at the microwave frequency. If this AC spin pumping signal were measured, the signal shall correspond to the inverse transverse spin Hall effect.

The longitudinal spin Hall effect in a ferromagnet has also been studied by non-local spin transport via Cu or YIG. In a Py/Cu/Py non-local spin valve, as shown in the inset of Fig. 10(a), Qin *et al.*

[58] observed an asymmetric non-local resistance. The asymmetric signal as illustrated in Fig. 10(c), arises from the longitudinal spin Hall effect in Py2, where the spin current is injected nonlocally from Py1. The reverse effect shown in Fig. 10(b) also exhibits similar behavior. The effective $\theta_{||}$ (denoted as $\alpha_{SH}$ in their paper) is coupled with the spin diffusion length $\lambda_{Py}$, where $\theta_{||} \lambda_{Py} = (0.041 \pm 0.010)$ nm at room temperature and $\theta_{||} \lambda_{Py} = (0.066 \pm 0.009)$ nm at 5 K. If we choose the spin diffusion length of Py to be about 3 nm [55], the extrapolated $\theta_{||}$ of Py is on the same order as those extrapolated by other methods as discussed above.

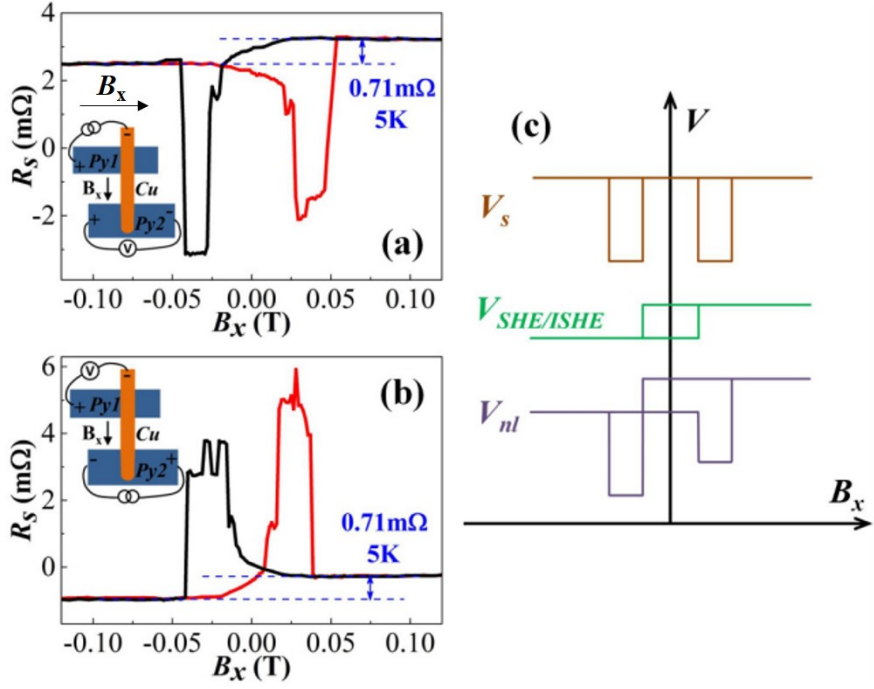

Figure 10 Nonlocal spin valve for measuring the charge-spin conversion via (a) the inverse longitudinal spin Hall effect and (b) the longitudinal spin Hall effect. (c) Sketches of the expected data. The conventional nonlocal magnetoresistance $V_s$ is symmetric about the y-axis, while the signal $V_{SHE/ISHE}$ due to the charge-spin conversion is resembles the magnetization hysteresis. The total voltage $V_{nl}$ is the superposition of $V_s$ and $V_{SHE/ISHE}$. Images adapted from Ref. [58].

Since the anomalous Hall effect and the longitudinal spin Hall effect may have similar origins, one may speculate that there are quantitative correlations between the two. Omori *et al.* [59] studied the temperature-dependent anomalous Hall resistivity and longitudinal spin Hall resistivity with a non-local spin valve structure similar to that in Ref. [58]. As shown in Fig. 11, the extrapolated longitudinal spin Hall resistivity ($\rho_{xy}^{SHE} = \theta_{||}\rho$, where $\rho$ is the electric conductivity) exhibits much stronger temperature dependence than the anomalous Hall resistivity ($\rho_{xy}^{AHE}$). It is argued that in the skew scattering mechanism, the longitudinal spin Hall resistivity may scale with the anomalous Hall resistivity by the spin polarization [59]. But such a relation may not hold for other mechanisms such as intrinsic mechanism and side jump.

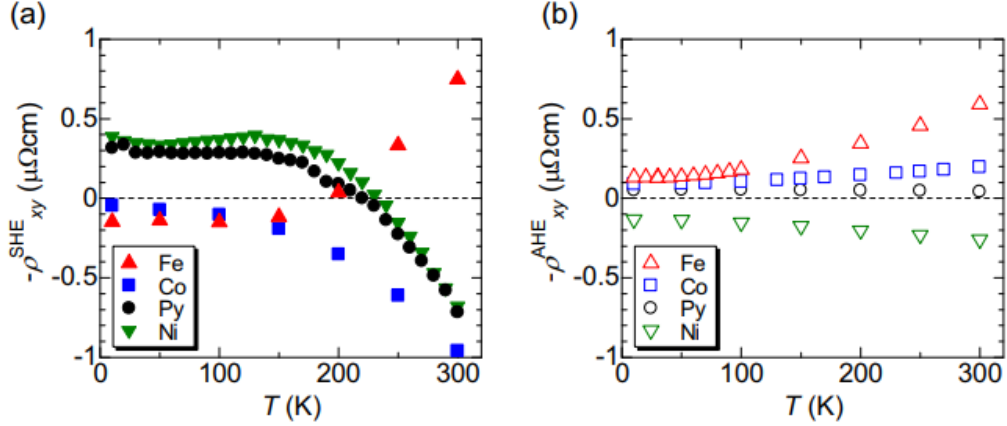

Figure 11 The temperature dependence of (a) longitudinal spin Hall resistivity and (b) anomalous Hall resistivity in Py, Fe, Co and Ni. Images adapted from Ref. [59].

The longitudinal spin Hall effect in a ferromagnet can be used to generate out-of-plane polarized spin current, an important application that is difficult to realize in nonmagnetic materials. As shown in Fig. 12(a), Gibbons *et al.* [60] demonstrated the feasibility by using a spin valve structure IrMn/FeGd/Hf/CoFeB, where FeGd and CoFeB are two magnetic layers. By measuring the second harmonic Hall signal, the spin-orbit torques acting on the CoFeB magnetization is deduced. A portion of the spin-orbit torque is attrbibuted to the field-like torque due to a spin current with spin direction parallel to $(\mathbf{y} \cdot \mathbf{m}_{FeGd})\mathbf{m}_{FeGd}$ (consistent with the longitudinal spin Hall effect described by the first term in Eq. (1)) where $\mathbf{y}$ is the in-plane direction perpendicular to the applied charge current, and $\mathbf{m}_{FeGd}$ is a unit vector along the FeGd magnetization direction. The extrapolated field-like spin-torque efficiency (similar to that of a spin Hall angle) is $0.009 \pm 0.002$ for FeGd.

Iihama *et al.* [61] demonstrated the spin-transfer torque induced by the longitudinal spin Hall effect by measuring the damping enhancement/suppression in a CoFeB/Cu/Py trilayer. The sample structure is shown in Fig. 12(b), where an in-plane charge current generates a spin current with spin direction parallel with the CoFeB magnetization via the longitudinal spin Hall effect. The spin current generates an anti-damping torque that enhances or reduces the damping of the Py layer depending on the electric current and Py magnetization directions, as shown in Fig. 12(b). The effective damping-like spin-torque efficiency extracted for the CoFeB layer is as large as $0.14 \pm 0.05$ with the same sign as that of Ta.

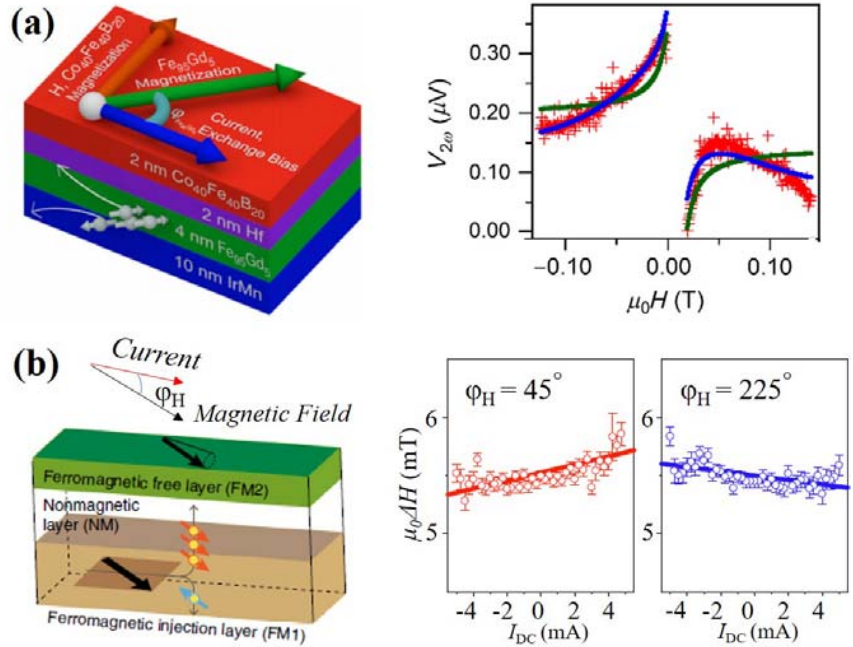

Figure 12 Longitudinal spin Hall effect-enabled spin-orbit torque studied in (a) a FeGd/Hf/CoFeB spin valve structure, via second harmonic planar Hall measurement, and (b) a Py/Cu/CoFeB spin valve structure, via damping enhancement/reduction measurement. Images adapted from Refs. [60] and [61].

It is worth pointing out that in both experimental demonstrations [60, 61], the spin polarization was assumed to be aligned in the same direction as the magnetization of the ferromagnetic spin current generator. This assumption neglects the transverse spin Hall effect, which allows the ferromagnet to generate spin current with spin polarization transverse to the magnetization. Taking into consideration of the transverse spin Hall effect may impact the analysis of these experimental results in a quantitative way.

## 7. Experimental Observation of Conversion between Charge Currents and Transverse Spins

The study of conversion between charge current and transversely polarized spin current requires the spin current of interest to have a spin direction perpendicular to the magnetization of a ferromagnet, therefore the sample under study is usually a spin valve with non-collinear magnetization configurations.

Tian *et al.* [62] investigated the spin Seebeck effect in a YIG/Cu/Co trilayer, where the magnetization of YIG and Co are decoupled. As shown in Fig. 13(a), for low applied magnetic field, it is possible to realize arbitrary angles between the YIG and Co magnetization. The thermal voltage signal consists of the anomalous Nernst effect from Co itself, $V_{ANE}$, and $\Delta V_{SSE}$, which is due to the inverse longitudinal(transverse) spin Hall effect if the YIG and Co magnetizations are aligned parallel(perpendicular) with each other. $V_{ANE}$ and $\Delta V_{SSE}$ can be separated because the two signals have different dependences on the switching of Co and YIG magnetizations, respectively. As shown in Fig. 13(b), the extracted $\Delta V_{SSE}$ appears to be independent of the relative angle between Co and YIG magnetization. This result not only shows that Co can convert transversely polarized spin current into an in-plane voltage, but also implies that the effective transverse spin Hall coefficient $\theta_\perp$ may be close to the longitudinal spin Hall coefficient $\theta_\parallel$ in their system.

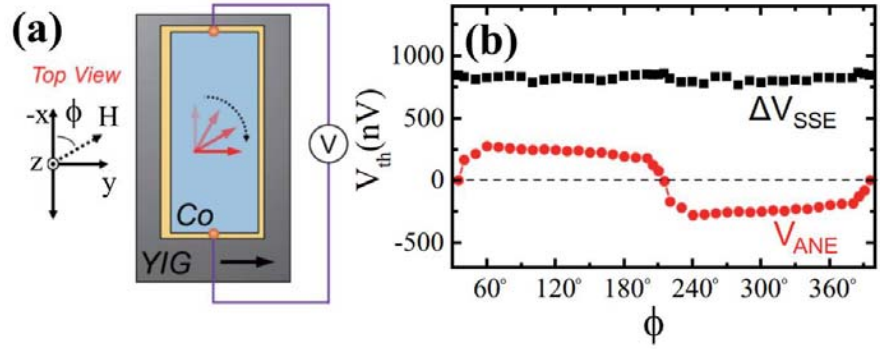

Figure 13 (a) Measurement configuration to study the spin-charge conversion in Co as a function of magnetization of Co. A temperature gradient is applied normal to the films. (b) The ANE of Co is found to reverse when Co magnetization reverses, while the spin-charge conversion signal due to spin current injected from YIG to Co is found to be independent of Co magnetization direction. Images adapted from Ref. [62].

As shown in Fig. 14(a), Das *et al.* [63] carried out nonlocal spin transport measurements on a Py/YIG/Py structure, where the spin transport is mediated by magnons in YIG. The exchange coupling between Py and YIG magnetizations were shown to be weak. As illustrated in Fig. 14(b), the Py magnetization can be aligned perpendicular to YIG magnetization under low magnetic field (5mT), where spin current generation and detection are realized by transverse spin Hall effect and inverse transverse spin Hall effect. When magnetic field is large (200 mT), the Py magnetization is aligned parallel to YIG magnetization, where spin current generation and detection are realized by longitudinal spin Hall effect and inverse longitudinal spin Hall effect. Shown in Fig. 14(c), the signals measured in the transverse spin hall configuration (pink region) is 50% that measured in the longitudinal spin Hall configuration (blue region), indicating $\theta_\perp$ is lower than $\theta_\parallel$ in their samples. Similar behavior was also observed in $Co_{60}Fe_{20}B_{20}$, in a similar device [64].

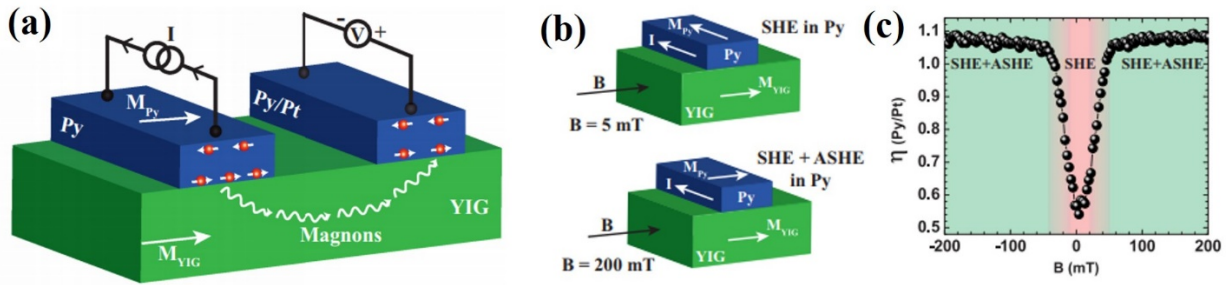

Figure 14 (a) Measurement setup for detecting the charge-spin conversion in Py via magnon mediated spin transport through YIG. (b) Transverse configuration at low magnetic field and longitudinal configuration at high magnetic field. (c) Transverse (low field data) and Longitudinal (high field data) spin Hall coefficient of Py scaled with the spin Hall angle of Pt. Images adapted from Ref. [63].

It should be pointed out that in Ref. [63], the transverse spin Hall effect was referred to as a magnetization-independent spin Hall effect, while the longitudinal spin Hall effect was attributed to a superposition between the magnetization-independent spin Hall effect and an anomalous spin Hall effect (ASHE) related to the anomalous Hall effect. This is mathematically equivalent to our description using the longitudinal and transverse spin Hall effects, as can be understood from Eq. (1). If the magnetization is in the film plane with an angle $\varphi$ from the electric field, the first two terms in Eq. (1) can be rewritten as

$\sigma_{\parallel}[\mathbf{m} \cdot (\hat{\mathbf{z}} \times \mathbf{E})]\mathbf{m} + \sigma_{\perp}\mathbf{m} \times [(\hat{\mathbf{z}} \times \mathbf{E}) \times \mathbf{m}] = \sigma_{\perp}(\hat{\mathbf{z}} \times \mathbf{E}) + (\sigma_{\parallel} - \sigma_{\perp})[\mathbf{m} \cdot (\hat{\mathbf{z}} \times \mathbf{E})]\mathbf{m}$ , where $\sigma_{\perp}$ corresponds to the magnetization-independent spin Hall conductivity $\sigma_{SHE}$, and $\sigma_{\parallel} - \sigma_{\perp}$ corresponds to the magnetization-dependent spin anomalous Hall conductivity $\sigma_{ASHE}$, as described in Ref. [63].

While transversely polarized spin current can be generated from a ferromagnet and influence a neighboring layer, it also can influence the ferromagnet itself. Wang *et al.* [65] showed that in a single-layer ferromagnet, the internally generated transversely polarized spin current results in equal and opposite spin torques at the surfaces of the ferromagnet. Due to the analogy to the anomalous Hall effect, as shown in Fig. 15(a-b), the current-induced spin torque within a ferromagnet is termed as the anomalous spin-orbit torque. The anomalous spin-orbit torque is measured by the magneto-optic-Kerr effect (MOKE). Due to finite penetration depth of light in Py films, the MOKE response from the top and bottom surfaces will not fully cancel out, leading to a net signal as shown in Fig. 15(c). The strength of the anomalous spin-orbit torque does not vary with different interfaces, suggesting it arises from a bulk-generated transversely polarized spin current. The effective transverse spin Hall angle, $\theta_{\perp}$ is extrapolated to be about 0.05, comparable to the spin Hall angle of Pt [9]. Similar effects are also observed in Co, Fe and Ni, suggesting the anomalous spin-orbit torque is a universal phenomenon for all ferromagnetic conductors.

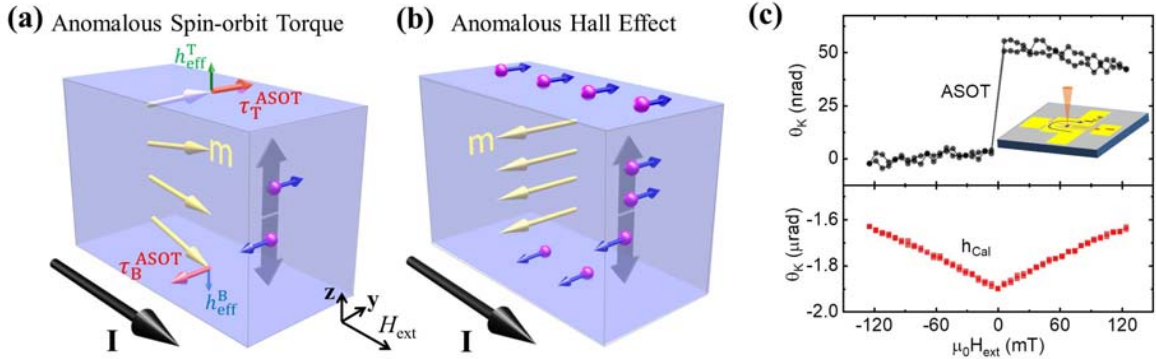

Figure 15 (a) Illustration of the anomalous spin-orbit torque when current is applied parallel with magnetization. The spin torques $\tau_T^{ASOT}$ and $\tau_B^{ASOT}$ are in the $y$ direction, which are equivalent to effective fields $h_{eff}^T$ and $h_{eff}^B$ in the $z$ direction. Here blue arrows on purple spheres represent electron spin directions and grey arrows represent spin flow directions. (b) Illustration of the anomalous Hall effect when current is applied perpendicular with magnetization. Due to the imbalance of majority and minority electrons, a net voltage accumulation is between the top and bottom surfaces. (c) Exemplary data of the MOKE signals due to the anomalous spin-orbit torque and an out-of-plane calibration field ($h_{Cal}$). The inset shows the experimental structure. Images adapted from Ref. [65].

Humphries *et al.* [66] reported the first observation of the transverse spin Hall effect with spin rotation. The samples studied have a spin-valve structure of PML/Cu/Py, where PML is a perpendicularly magnetized layer and Py has an in-plane magnetization. When an in-plane electric field **E** is applied, as shown in Fig. 16(a), the transverse spin Hall effect with spin rotation in PML leads to an out-of-plane flowing spin current $\mathbf{Q}_z^R$ with spin direction parallel to $\mathbf{m} \times (\mathbf{E} \times \mathbf{z})$, where **m** is the PML magnetization, and **z** is the out-of-plane direction. The spin current flows through Cu and exerts a spin torque on Py, which is probed by magento-optic-Kerr-effect (MOKE) [67]. The spin current reverses polarization when the PML magnetization switches, resulting in a reversed spin-orbit torque on Py magnetization, as shown in Fig. 16(b). The effective coefficient $\theta_\perp^R$ for the transverse spin Hall effect with spin rotation was extrapolated to be $(4.8 \pm 0.6) \times 10^{-3}$. The reverse effect – the inverse transverse spin Hall effect with spin rotation – was also detected in the similar structure [68], as shown in Fig. 16(c-d). When a perpendicular temperature gradient is applied, the spin Seebeck effect in Py drives a spin current with spin direction parallel to Py magnetization $\mathbf{m}_{Py}$. This spin current generates a voltage in the direction $(\mathbf{m} \times \mathbf{m}_{Py}) \times \mathbf{z}$ via the inverse transverse spin Hall effect with spin rotation of PML. The thermal voltage signal $V_{TH}$ resembles the magnetization hysteresis of Py and reverses as PML magnetization reverses as observed in Fig. 16(d).

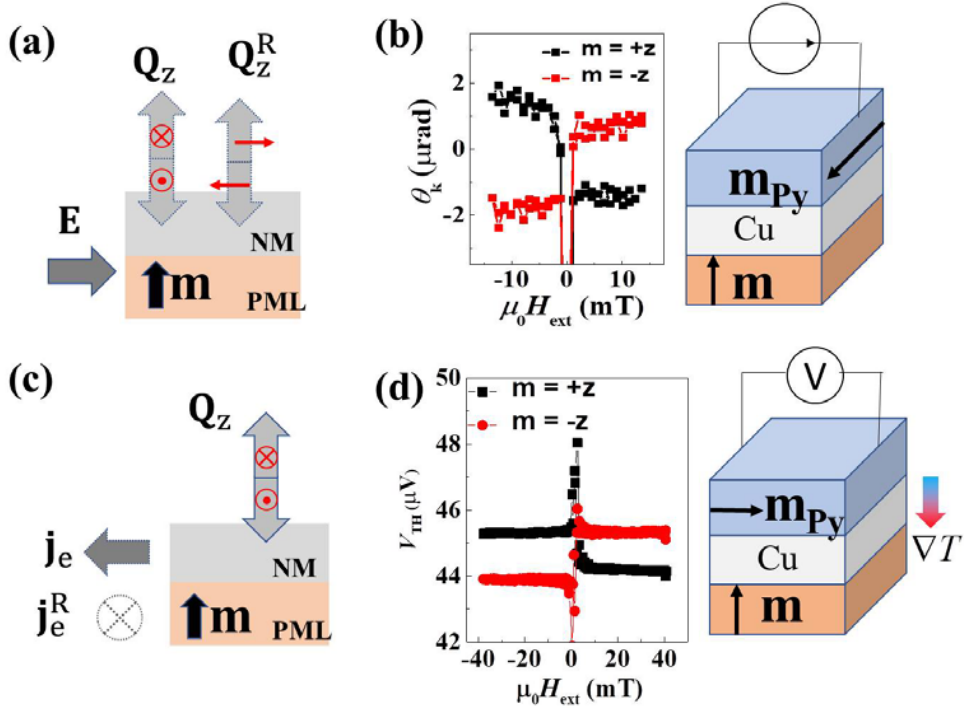

Figure 16 (a) Illustration of current-induced spin current with conventional spin Hall symmetry, $\mathbf{Q}_z$ and spin current with spin rotation, $\mathbf{Q}_z^R = \mathbf{Q}_z \times \mathbf{m}$. (b) Polar MOKE signal $\Psi_{polar}$ that reveals the torque generated by $\mathbf{Q}_z^R$, which reverses as the PML magnetization reverses. (c) Illustration of spin current-driven charge current due to inverse transverse spin Hall effect, $\mathbf{j}_e$, and inverse transverse spin Hall effect with spin rotation, $j_e^R$. (d) Voltage generated by the spin Seebeck effect with spin rotation, which is odd with the PML magnetization. Images adapted from Ref. [66] and [68].

Transverse spin Hall effect with spin rotation allows the generation of out-of-plane polarized spin current from an in-plane magnetized film. The out-of-plane polarized spin current can switch a perpendicular magnetization via anti-damping process, which is recently demonstrated by Baek *et al.* [69]. Illustrated in Fig. 17(a), the sample studied is FM/Ti/CoFeB, where the CoFeB is magnetized out of film plane, FM = Py, CoFeB is in-plane magnetized. They observed that the FM magnetized in the x-direction can generate a transversely polarized spin current with the spin direction in the z-direction, characterized by the transverse spin Hall effect with spin rotation. The effective coefficients $\theta_\perp^R$ for CoFeB and Py are found to be $-0.014 \pm 0.001$ and $0.006 \pm 0.0006$, respectively. In addition, Baek *et al.* demonstrated that the out-of-plane polarized spin current can lead to a field-free magnetization switching of the perpendicularly magnetized CoFeB layer, as shown in Fig. 17(b).

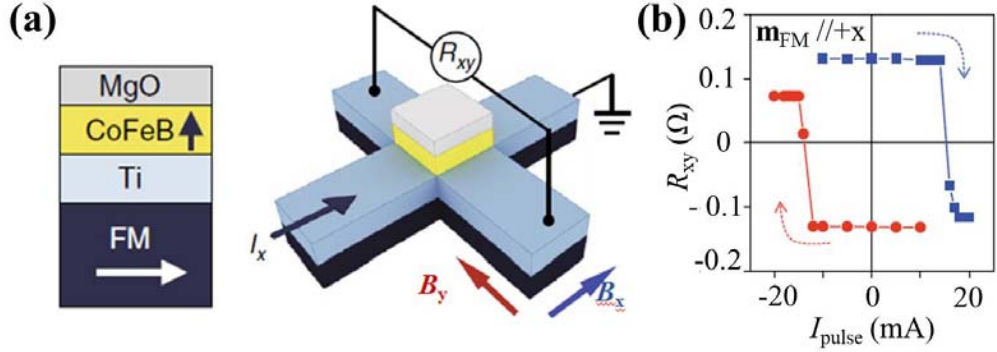

Figure 17 (a) Measurement geometry for field-free switching via out-of-plane polarized spin current. (b) Hall signal to show the reversal of CoFeB magnetization as a function of applied current pulse. Images adapted from Ref. [69].

Although both the longitudinal spin Hall effect with a tilted magnetization and the transverse spin Hall effect with spin rotation can be used toward generating out-of-plane polarized spin current, the latter may be practically advantageous because it is easier to fabricate an in-plane magnetized film than a film with magnetization partially tilted out of plane.

## 8. Other Charge-Spin Conversion in Ferromagnetic Materials

Besides the longitudinal and transverse spin Hall effects, a ferromagnetic material with a tilted magnetization can also possess a charge-spin conversion with planar Hall symmetry [32, 70]. Safranski *et al.* [71] reported the observation of planar Hall torque in a ferromagnet/nonmagnet multilayer. Shown in Fig. 18 (a), when the magnetization is tilted in the *xz*-plane, an electric current in the *x*-direction can generate an out-of-plane flowing spin current with spin direction parallel with the magnetization, due to the planar Hall effect. Such a spin current can apply a spin torque back on the magnetization, modifying the magnetic damping. Unlike the spin Hall torque, which scales with $m_y$, the planar Hall torque scales with $m_x m_z$, as shown in Fig. 18, following the planar Hall symmetry.

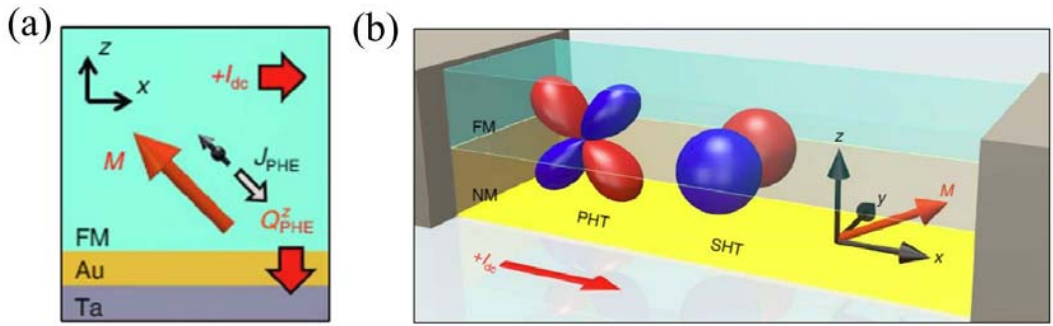

Figure 18 Current-induced spin-orbit torque with planar Hall symmetry. Images adapted from Ref. [71].

The examples we have reviewed above are focused on ferromagnets. But the same symmetry argument also applies to ferrimagnet and non-collinear antiferromagnetic materials. The magnetic order of non-collinear antiferromagnets such as $Mn_3Ir$ and $Mn_3Sn$ breaks symmetries just like the magnetization in a ferromagnet. Accordingly, a large anomalous Hall voltage signal has been predicted [72] and observed [73] in non-collinear antiferromagnets. Recently, Kimata *et al.* reported the observation of magnetic spin Hall effect in $Mn_3Sn$ [74], which is an antiferromagnetic analogy to the transverse spin Hall effect with spin rotation in ferromagnets.

## 9. Outlook

From a symmetry analysis, we have shown that a simple ferromagnetic conductor possesses a complicated spin current conductivity, composed of a longitudinal spin Hall effect, transverse spin Hall effect, transverse spin Hall effect with spin rotation, together with their inverse effects. These effects have been experimentally demonstrated and theoretically formulated in recent years. However, there remain many unanswered questions: (1) What are the microscopic mechanisms that give rise to the charge-spin conversion in the ferromagnetic metal: is the longitudinal spin-charge conversion solely a bulk effect and sharing a same microscopic origin as the anomalous Hall effect? What governs the transverse spin Hall effect, interface or bulk? What type of material engineering can enhance or suppress these effects? (2) Like the studies on the spin Hall effect and inverse spin Hall effect in nonmagnetic materials, interface transparency plays a very important role. What is the appropriate model for the propagation of spin current within a ferromagnet, particularly for transversely polarized spin current? (3) Ferromagnetic conductors are ubiquitous in many fields of spin-orbitronics, but the transverse spin Hall effect, which generates transversely polarized spin current from the ferromagnet itself, were often neglected. Will the newly discovered spin-orbit effects in ferromagnets challenge previous understandings of spin-orbit effects, such as spin-orbit torque and spin pumping-spin galvanic effect [5]?

The spin-charge conversion from ferromagnetic conductors also provides new opportunities. Traditionally nonmagnetic materials are often used as a spin current source. Limited by the geometry and the symmetry of the spin Hall effect, the spin current generated that flows out of a thin film is only polarized in-plane for most materials. This limitation is lifted by the ferromagnet, from which the polarization of the generated spin current can be manipulated by the magnetization. The generation of out-of-plane polarized spin current may enable new device designs in magnetic memories, domain wall and skyrmion manipulation and magnetic nano-oscillators.

The investigation of charge-spin conversion in nonmagnetic materials have led to the discovery of new transport behaviors such as the spin Hall magnetoresistance [75, 76] and unidirectional spin Hall magnetoresistance [77]. Very recently, the anomalous Hall magnetoresistance [78] has also been discovered, which is directly correlated with the longitudinal spin Hall effect of the ferromagnet. We expect that a comprehensive understanding of charge-spin conversion in ferromagnetic conductors will lead to the discovery of more unique transport behaviors.

**Acknowledgement**

The work done at the University of Denver is supported by the Faculty Research Fund. V.P.A. acknowledges support under the Cooperative Research Agreement between the University of Maryland and the National Institute of Standards and Technology Center for Nanoscale Science and Technology, Award 70NANB14H209, through the University of Maryland. We would also like to thank Mark Stiles and Emilie Jue for critical reading of the manuscript.